\documentclass[12pt]{article}

\usepackage[totalwidth=460truept,totalheight=625truept]{geometry}
\usepackage{amsfonts,latexsym,amssymb,amsmath,graphicx,accents,slashed,subfigure}
\usepackage[T1]{fontenc}
\usepackage[hidelinks]{hyperref}

\global\arraycolsep=1truept
\numberwithin{equation}{section}

\begin{document}

\null

\bigskip

\begin{center}
{\huge \textbf{Fakeons,\ Quantum Gravity And}}

\vskip.5truecm

{\huge \textbf{The Correspondence Principle}}

\vskip1truecm

\textsl{Damiano Anselmi}

\vskip .1truecm

\textit{Dipartimento di Fisica \textquotedblleft Enrico
Fermi\textquotedblright , Universit\`{a} di Pisa}

\textit{Largo B. Pontecorvo 3, 56127 Pisa, Italy,}

\textit{and INFN, Sezione di Pisa,}

\textit{Largo B. Pontecorvo 3, 56127 Pisa, Italy,}

damiano.anselmi@unipi.it

\vskip1truecm

\textbf{Abstract}
\end{center}

The correspondence principle made of unitarity, locality and
renormalizability has been very successful in quantum field theory. Among
the other things, it helped us build the standard model. However, it also
showed important limitations. For example, it failed to restrict the gauge
group and the matter sector in a powerful way. After discussing its
effectiveness, we upgrade it to make room for quantum gravity. The unitarity
assumption is better understood, since it allows for the presence of
physical particles as well as fake particles (fakeons). The locality
assumption is applied to an interim classical action, since the true
classical action is nonlocal and emerges from the quantization and a later
process of classicization. The renormalizability assumption is refined to
single out the special role of the gauge couplings. We show that the
upgraded principle leads to an essentially unique theory of quantum gravity.
In particular, in four dimensions, a fakeon of spin 2, together with a
scalar field, is able to make the theory renormalizable while preserving
unitarity. We offer an overview of quantum field theories of particles and
fakeons in various dimensions, with and without gravity.

\vskip 1truecm

Proceedings of the conference \textquotedblleft \textit{Progress and Visions
in Quantum Theory in View of Gravity: Bridging foundations of physics and
mathematics}\textquotedblright , Max Planck Institute for Mathematics in the
Sciences, Leipzig, October 2018

\vfill\eject

\section{Introduction}

\setcounter{equation}{0}

According to Bohr's correspondence principle, it must be possible to obtain
the laws of classical mechanics from those of quantum mechanics in the limit
of large quantum numbers, or, more generally, the \textit{classical limit}.
We can view the principle as a guideline for the selection of theories.
Behind the necessity of such a selection is the fact that our observational
power is considerably reduced when we explore the microscopic world.

At the classical level, we can uncover the physical laws relatively easily,
because we can make a large number of experimental observations at the same
time without disturbing the system, at least in principle. On the contrary,
at the quantum level, our possibilities of observing the microscopic world
are limited by several factors, including the laws of physics themselves,
e.g. the uncertainty principle, as well as our physical constitution, the
dimensions of the cells and atoms of which we are made. The human beings are
clumps of atoms that are trying to \textquotedblleft
understand\textquotedblright \footnote{%
When we \textquotedblleft understand\textquotedblright\ something previously
unknown to us, we just establish analogies, relations, correspondences with
phenomena that are more familiar to us. Ultimately, \textquotedblleft
understanding\textquotedblright\ just means \textquotedblleft getting used
to\textquotedblright .} the laws that govern scales of magnitude that are
billions of billions of times smaller than the smallest ingredient they are
made of. Very probably, this is a vicious circle. Below certain scales of
magnitude the universe may become unknowable to us.

Our thought is, so to speak, \textquotedblleft classical\textquotedblright ,
because it is shaped by the interactions between us and the classical
environment where we live. The fundamental concepts of our logic (such as
existence, origin, time, space, cause, effect, principle, consequence, etc.)
are inherited from that environment. Unlikely they are absolute. They might
just be useful approximations, or effective descriptions with limited ranges
of applicability.

For these reasons, a quantum theory is not built from scratch, but instead
guessed from another theory that is more familiar to us, which we call
classical and which is later quantized. Unless there is a sort of
correspondence between the two, our chances of understanding the quantum
world are minimal.

Quantum mechanics forced us to waive determinism, which we used to take for
granted. The lesson is that at any point we may have to modify the laws of
physics and even the basic principles of our thinking in a profound way.
Instead of assuming that our knowledge is \textquotedblleft
universal\textquotedblright\ and extends straightforwardly to the unknown
portion of the universe, we must admit that the \textquotedblleft
principles\textquotedblright\ suggested by our classical experiences are
just temporary work hypotheses.

With the help of the devices we build, we can extend the exploration of the
world way beyond the limits of our direct perception. However, the devices
have limits as well, because they are macroscopic, like us. When we build
them, we tacitly assume that the laws of nature derived from the observation
of the known portion of the world remain valid when we explore the unknown
portion. If everything works as expected, the assumption is validated. Yet,
this does not prove that those laws are universal, i.e. hold for arbitrarily
short time intervals, or arbitrarily large energies.

It is reasonable to expect that, when the energies we explore are not too
high, the laws of nature remain to some extent similar to the laws obeyed by
the phenomena occurring in the classical environment that surrounds us.
Bohr's correspondence principle codifies such a similarity up to the atomic
distances, which are the realm of quantum mechanics. What happens when we
explore smaller distances? Conceivably, the correspondence will become
weaker and weaker and our instruments will not help us indefinitely.

The first descent to smaller distances is quantum field theory. There, we
talk about classical limit in a different sense, which may not even refer to
true classical phenomena, as quantum chromodynamics shows. Yet, an upgraded
version of the correspondence principle does emerge, summarized by the
requirements of unitarity, locality and renormalizability. It leads us to
build the standard model of particle physics.

At the same time, crucial limitations appear. For example, we are still
unable to explain why the gauge group of the standard model is $U(1)\times
SU(2)\times SU(3)$. Moreover, the matter sector is only weakly constrained.
So far, all the attempts to unify the three interactions of nature encoded
in the standard model have failed. One possible explanation is that we have
not been clever enough, but another, sadder possibility is that the
correspondence between the macroscopic world we live in and the microscopic
world we wish to explore might be fading away. At some point all similarity
will eventually disappear and we will remain blind and powerless. For this
reason, we think that it is too risky to depart from the kind of
correspondence that has worked so far: we have to be as conservative as
possible.\ 

The second descent, quantum gravity, may require to reconsider or refine the
basic assumptions of the correspondence principle in nontrivial ways. In
this paper, we show that further upgrades are indeed available and lead to
an essentially unique solution to the problem of quantum gravity.

The basic new idea is the concept of fake particle (fakeon), which is able
to reconcile renormalizability and unitarity. In four dimensions, quantum
gravity is described by a triplet made by the graviton, a fakeon of spin 2
and a scalar field. The theory is very predictive and way more unique than
the standard model. We infer the upgraded correspondence principle from its
main properties and then give an overview of quantum field theories of
particles and fakeons in various dimensions, with and without gravity.

The fakeons are introduced by means of a novel quantization prescription 
\cite{LWgrav,fakeons} for the poles of the free propagators. In the
Euclidean region of the space $\mathcal{P}$ of the complexified external
momenta, a Feynman diagram is evaluated as usual, from the Euclidean version
of the theory. Elsewhere, it is evaluated from the Euclidean region by
analytic continuation up to the fakeon thresholds, which are the thresholds
associated with the processes involving fakeons. Above those thresholds the
diagram is evaluated by means of a nonanalytic operation, called average
continuation, which amounts to take the arithmetic average of the analytic
continuations that circumvent the threshold. Overall, we may view the
procedure as a nonanalytic Wick rotation \cite{LWformulation,LWunitarity}.
Finally, to have unitarity, the fakeons must be projected away from the
physical spectrum.

In a quantum field theory of particles and fakeons the quantization has to
be understood in a new way. To mention one thing, the starting classical
action, which is local, is just an interim action, because it is
unprojected, i.e. it contains the (classical counterparts of the) fakeons.
However, since the projection comes from the quantization, when we want to
reach the classical limit we must first quantize the theory and then
classicize it back. Only at the end of this procedure we obtain the true
classical action, which is nonlocal. We see that the locality assumption
must be understood anew and applied to the interim classical action.

In addition, the renormalizability requirement has to be formulated more
precisely, because the existing definitions do not make us appreciate the
peculiar role reserved to the gauge couplings. As far as the unitarity
requirement is concerned, it does not need to be modified, but it is
necessary to realize that it leaves room for both physical particles and
fakeons.

Among the other approaches to the problem of quantum gravity appeared in the
past decades, we mention string theory \cite{string}, loop quantum gravity 
\cite{loop}, holography \cite{ads} and asymptotic safety \cite{asafety}.
Most of them follow from correspondence principles that sound very \textit{%
ad hoc} and are based on assumptions that appear hard to justify. None of
them is as close to the standard model as the solution based on the fakeon
idea, which is a quantum field theory, admits a perturbative expansion in
terms of Feynman diagrams and allows us to make calculations with a
comparable effort (see refs. \cite{absograv,UVQG}).

String theory is criticized for being nonpredictive \cite{landscape}.
Moreover, its calculations often require mathematics that is not completely
understood. Loop quantum gravity is even more challenging, because it is at
an earlier stage of development. The AdS/CFT correspondence and the
asymptotic-safety program do not admit weakly coupled expansions. Our
solution bests its competitors in calculatibity, predictivity and
falsifiability. It is also rather rigid, because it contains only two new
parameters.

A theory of quantum gravity is supposed to shed light on a new understanding
of spacetime at the microscopic level. In the theories of particles and
fakeons, this is the violation of microcausality: at energies larger than
the fakeon masses, past, present and future lose meaning and there is no way
to tell the difference between cause and effect. From the theoretical and
experimental points of view, there is room for this prediction to be
accurate. The new physics is expected to emerge at energies around the
fakeon masses, which might be well below the Planck scale, possibly around 10%
$^{12}$GeV.

The paper is organized as follows. In section \ref{quanti} we recall the
fakeon idea. In section \ref{presco} we recall the formulation of quantum
gravity it leads to. In section \ref{predi} we study the dressed
propagators. In section \ref{class} we classicize quantum gravity. In
section \ref{upgra} we summarize the lessons learned and upgrade the
correspondence principle. Section \ref{conclusions} contains the conclusions.

\section{Fakeons}

\setcounter{equation}{0}\label{quanti}

In this section, we discuss the idea of fake particle introduced in ref. 
\cite{LWgrav}. We start from the crucial property, which is unitarity. Once
the $S$ matrix is written as $1+iT$, the unitarity equation $S^{\dag }S=1$
gives the optical theorem%
\begin{equation}
2\hspace{0.01in}\mathrm{Im}T=T^{\dag }T.  \label{optical}
\end{equation}%
A diagrammatic version of this identity is provided by the so-called cutting
equations \cite{cuttingeq}, which express the real part of a diagram as a
sum of \textquotedblleft cut diagrams\textquotedblright . The simplest
cutting equations are 
\begin{eqnarray}
2\hspace{0.01in}\text{Im}\left[ (-i)%
\raisebox{-1mm}{\scalebox{2}{$\rangle
\hspace{-0.075in}-\hspace{-0.07in}\langle$}}\,\right] =%
\raisebox{-1mm}{\scalebox{2}{$\rangle
\hspace{-0.075in}-\hspace{-0.14in}\slash\hspace{-0.015in}\langle$}} &=&\int 
\mathrm{d}\Pi _{f}\hspace{0.01in}\left\vert \raisebox{-1mm}{\scalebox{2}{$%
\rangle\hspace{-0.035in}-$}}\right\vert ^{2},  \label{cutd} \\
2\hspace{0.01in}\text{Im}\left[ (-i)\raisebox{-1mm}{\scalebox{2}{$-%
\hspace{-0.065in}\bigcirc\hspace{-0.065in}-$}}\right] =\hspace{0.01in}%
\raisebox{-1mm}{\scalebox{2}{$-\hspace{-0.065in}\bigcirc\hspace{-0.16in}%
\slash\hspace{0.015in}-$}} &=&\int \mathrm{d}\Pi _{f}\hspace{0.01in}%
\left\vert \raisebox{-1mm}{\scalebox{2}{$-\hspace{-0.035in}\langle$}}%
\right\vert ^{2},  \label{cutdd}
\end{eqnarray}%
where the integrals are over the phase spaces $\Pi _{f}$ of the final states 
\cite{peskin}.

Let $V$ denote the space of physical states. In various cases, it is
necessary to work with a larger space $W$, which contains also unphysical
states. The matrix element $\langle b|T|a\rangle $, where $|a\rangle
,|b\rangle \in W$, is given by the connected, amputated diagrams, with
external legs determined by $|a\rangle $ and $|b\rangle $. Under very
general assumptions, it is relatively easy to prove a diagrammatic identity
resembling (\ref{optical}) in $W$, which reads%
\begin{equation}
\frac{1}{2i}\left[ \langle b|T|a\rangle \hspace{0.01in}-\langle b|T^{\dagger
}|a\rangle \hspace{0.01in}\right] =\sum_{|n\rangle \in W}\langle b|T^{\dag
}|n\rangle (-1)^{\sigma _{n}}\langle n|T|a\rangle ,\qquad |a\rangle
,|b\rangle \in W,  \label{W}
\end{equation}%
where $\sigma _{n}$ can be 0 or 1, the unphysical states being those with $%
\sigma _{n}=1$. The identity (\ref{W}) is called pseudounitarity equation.

The cut propagators of the cut diagrams carry information about the
intermediate states $|n\rangle $ and the space $W$. Unitarity requires to
prove that the identity (\ref{W}) can be consistently projected onto $V$ to
give%
\begin{equation}
\frac{1}{2i}\left[ \langle b|T|a\rangle \hspace{0.01in}-\langle b|T^{\dagger
}|a\rangle \hspace{0.01in}\right] =\sum_{|n\rangle \in V}\langle b|T^{\dag
}|n\rangle \langle n|T|a\rangle ,\qquad |a\rangle ,|b\rangle \in V.
\label{V}
\end{equation}

What is crucial about the optical theorem is that it is not a linear
equation, but a quadratic one, so it mixes different orders of the loop
expansion. We can restrict the initial and final states $|a\rangle
,|b\rangle $ of (\ref{W}) at no cost, but it is not equally easy to restrict
the sum over $|n\rangle \in W$ to a sum over $|n\rangle \in V$. Thus, a
generic projection is inconsistent with unitarity: if we drop some states
from the set of initial and final states, they are generated back as
intermediate states $|n\rangle $ by the loop corrections. A projection that
is consistent with unitarity must be a very clever one.

These simple remarks show us that unitarity is an essentially loop property.
A tree-level action cannot be unitary \textit{per se}, because the
right-hand side of (\ref{cutdd}) is made of tree vertices only, but the
left-hand side is made of loops.

At present, the fakeon projection is the only example of consistent
projection, aside from the one that takes care of the Faddeev-Popov ghosts
and the temporal and longitudinal components of the gauge fields. As we
discuss below, the fakeon projection is actually very different from the
projection of gauge theories, to the extent that it leaves an important
remnant: the violation of microcausality.

Consider the propagator%
\begin{equation}
G(p,m)=\frac{1}{p^{2}-m^{2}}.  \label{prop}
\end{equation}%
Endowed with the Feynman prescription ($p^{2}\rightarrow p^{2}+i\epsilon $),
it becomes%
\begin{equation}
G_{+}(p,m,\epsilon )=\frac{1}{p^{2}-m^{2}+i\epsilon }  \label{propF}
\end{equation}%
and describes a particle of mass $m$. Consider the identity (\ref{cutd}),
with vertices equal to $-i$. If $P$ denotes the propagator of the
intermediate line on the left-hand side, (\ref{cutd}) gives the inequality $%
\mathrm{Im}[-P]\geqslant 0$. Taking $P=G_{+}$, we get 
\begin{equation}
\mathrm{Im}\left[ -\frac{1}{p^{2}-m^{2}+i\epsilon }\right] =\pi \delta
(p^{2}-m^{2}),  \label{opti}
\end{equation}%
which is indeed nonegative.

If we multiply (\ref{propF}) by a minus sign, we obtain a ghost, since $%
P=-G_{+}$ satisfies $\mathrm{Im}[-P]\leqslant 0$, in contradiction with the
optical theorem. However, if we also replace $+i\epsilon $ with $-i\epsilon $%
, the right-hand side of (\ref{opti}) does not change and the optical
theorem remains valid. The moral of the story is that we can in principle
have both propagators%
\begin{equation*}
G_{\pm }(p,m,\epsilon )=\pm \frac{1}{p^{2}-m^{2}\pm i\epsilon },
\end{equation*}%
since both fulfill the identity (\ref{cutd}).

However, if we integrate directly on Minkowski spacetime, the presence of
both $G_{+}$ and $G_{-}$ in the same Feynman diagram originates nonlocal
divergences at $\epsilon \neq 0$ \cite{ugo} and worse problems for $\epsilon
\rightarrow 0$. If, on the other hand, we start from the Euclidean version
of the theory, we find that the Wick rotation is not analytic and must be
defined anew \cite{LWformulation,LWunitarity}. One way to uncover the
concept of fake particle is precisely to make the Wick rotation work in a
way that is compatible with unitarity.

Let us multiply (\ref{prop}) by $\pm $, to emphasize that what we are going
to say applies irrespectively of the sign of the residue. Following \cite%
{LWgrav}, write 
\begin{equation*}
\pm \frac{p^{2}-m^{2}}{(p^{2}-m^{2})^{2}}
\end{equation*}%
and eliminate the singularity by introducing an infinitesimal width $%
\mathcal{E}$, to define the fakeon propagator as 
\begin{equation}
\mathbb{G}_{\pm }(p,m,\mathcal{E}^{2})=\pm \frac{p^{2}-m^{2}}{%
(p^{2}-m^{2})^{2}+\mathcal{E}^{4}}=\pm \frac{1}{2}\left[ G_{+}(p,m,\mathcal{E%
}^{2})-G_{-}(p,m,\mathcal{E}^{2})\right] .  \label{peps}
\end{equation}%
The propagator $\mathbb{G}_{+}(p,m,\mathcal{E}^{2})$ describes a \textit{%
fakeon plus}, while the propagator $\mathbb{G}_{-}(p,m,\mathcal{E}^{2})$
describes a \textit{fakeon minus}. Note that $\mathbb{G}_{\pm }(p,m,\mathcal{%
E}^{2})$ vanishes on shell at $\mathcal{E}>0$. This suggests that it does
not truly propagate a particle.

Formulas (\ref{peps}) are not the end of the story: we still have to explain
how to use them inside the Feynman diagrams. The matter is technically
involved, but a shortcut, called \textit{average continuation }\cite%
{LWformulation,fakeons}, allows us to jump directly to the final result.

Let $\mathcal{P}$ denote the hyperplane of the complexified external
momenta. In the Euclidean region we can evaluate the diagram from the
Euclidean version of the theory. No particular prescription or attention is
needed there. The result can be analytically continued within $\mathcal{P}$
up to the \textquotedblleft fakeon thresholds\textquotedblright , i.e. the
thresholds associated with the \textquotedblleft would-be
processes\textquotedblright\ involving fakeons. The fakeon thresholds are
overcome by means of the average continuation, i.e. by taking the arithmetic
average of the analytic continuations that circumvent the threshold. At the
end, the probabilities of the processes that involve the fakeons vanish.
This makes it possible to project the fakeons away and have unitarity.

Analyticity no longer holds in the usual sense. It is replaced by \textit{%
regionwise analyticity}. For every Feynman diagram, $\mathcal{P}$ is divided
into disjoint regions. In each region the diagram evaluates to an analytic
function. The main region is the Euclidean one, where the Wick rotation is
analytic. The other regions can be reached unambiguously from the Euclidean
one by means of the average continuation.

\section{Quantum gravity}

\label{presco}

The interim classical action of quantum gravity coupled to matter can be
basically written in two ways. If we use higher derivatives, it reads%
\begin{equation}
S_{\text{QG}}=-\frac{1}{2\kappa ^{2}}\int \sqrt{-g}\left[ 2\Lambda
_{C}+\zeta R+\alpha \left( R_{\mu \nu }R^{\mu \nu }-\frac{1}{3}R^{2}\right) -%
\frac{\xi }{6}R^{2}\right] +S_{\mathfrak{m}}(g,\Phi ),  \label{SQG}
\end{equation}%
where $\Phi $ are the matter fields and $S_{\mathfrak{m}}$ is the
covariantized action of the standard model, or an extension of it, once we
equip it with the nonminimal couplings that are required by renormalization.
The reduced Planck mass is $\bar{M}_{\text{Pl}}=M_{\text{Pl}}/\sqrt{8\pi }=%
\sqrt{\zeta }/\kappa $ and $\alpha $, $\xi $, $\zeta $ and $\kappa $ are
real positive constants, while $\Lambda _{C}$ can be positive or negative.
Here and below the integration measure $\mathrm{d}^{4}x$ is understood.

The second way is obtained by adding extra fields to eliminate the higher
derivatives. The result is, at $\Lambda _{C}=0$, 
\begin{equation}
\mathcal{S}_{\text{QG}}(g,\phi ,\chi ,\Phi )=S_{\text{H}}(g)+S_{\chi
}(g,\chi )+S_{\phi }(\tilde{g},\phi )+S_{\mathfrak{m}}(\tilde{g}\mathrm{e}%
^{\kappa \phi },\Phi ),  \label{SQG2}
\end{equation}%
where~$\tilde{g}_{\mu \nu }=g_{\mu \nu }+2\chi _{\mu \nu }$ and 
\begin{eqnarray}
&&S_{\text{H}}(g)=-\frac{\zeta }{2\kappa ^{2}}\int \sqrt{-g}R,\qquad S_{\phi
}(g,\phi )=\frac{3\zeta }{4}\int \sqrt{-g}\left[ \nabla _{\mu }\phi \nabla
^{\mu }\phi -\frac{m_{\phi }^{2}}{\kappa ^{2}}\left( 1-\mathrm{e}^{\kappa
\phi }\right) ^{2}\right] ,  \notag \\
&&S_{\chi }(g,\chi )=S_{\text{H}}(\tilde{g})-S_{\text{H}}(g)-2\int \chi
_{\mu \nu }\frac{\delta S_{\text{H}}(\tilde{g})}{\delta g_{\mu \nu }}+\frac{%
\zeta ^{2}}{2\alpha \kappa ^{2}}\int \left. \sqrt{-g}(\chi _{\mu \nu }\chi
^{\mu \nu }-\chi ^{2})\right\vert _{g\rightarrow \tilde{g}}.  \label{ss}
\end{eqnarray}%
The expression of (\ref{SQG2}) for $\Lambda _{C}\neq 0$ can be found in ref. 
\cite{absograv}.

In addition to the matter fields $\Phi $, the theory describes the graviton,
a scalar $\phi $ of squared mass $m_{\phi }^{2}=\zeta /\xi $\ and a spin-2
field $\chi _{\mu \nu }$ of squared mass $m_{\chi }^{2}=\zeta /\alpha $.
Making formula (\ref{ss}) more explicit, it is easy to show that the $\chi
_{\mu \nu }$ quadratic action is a covariantized Pauli-Fierz action with the
wrong overall sign, plus some nonminimal terms \cite{absograv}. This means
that, to have unitarity, the field $\chi _{\mu \nu }$ must be quantized as a
fakeon. Instead, the $\phi $ action has the correct sign, so $\phi $ can be
quantized either as a fakeon or a physical particle. Depending on which
option we choose, we have a graviton/fakeon/fakeon (GFF) theory or a
graviton/scalar/fakeon (GSF) theory.

Formula (\ref{SQG2}) shows that the matter fields $\Phi $ are sensitive to
the whole triplet $\{g_{\mu \nu },\chi _{\mu \nu },\allowbreak\phi \}$ of
quantum gravity, through the modified metric $\tilde{g}_{\mu \nu }\mathrm{e}%
^{\kappa \phi }=(g_{\mu \nu }+2\chi _{\mu \nu })\mathrm{e}^{\kappa \phi }$.
This means that the usual vertices that couple matter to gravity are
accompanied by similar vertices that couple matter to $\chi _{\mu \nu }$ and 
$\phi $. The theory predicts modified gravity-matter couplings. In
particular, the effective graviton-matter vertices receive loop corrections
due to the exchanges of $\chi _{\mu \nu }$ and $\phi $, similar to the QED
corrections studied in ref. \cite{matter}.

Renormalizability can be straightforwardly proved from the action (\ref{SQG}%
), because it does not depend on the quantization prescription \cite{fakeons}%
. Therefore, the beta functions coincide with those of the Stelle theory,
which is the theory obtained by quantizing all the degrees of freedom by
means of the usual Feynman prescription \cite{stelle}. In the Stelle theory $%
\chi _{\mu \nu }$ is a ghost instead of a fakeon and unitarity is violated
at energies larger than $m_{\chi }$.

\section{The dressed propagators}

\setcounter{equation}{0}\label{predi}

Thanks to the average continuation, calculating loop diagrams with the
fakeon prescription does not require much more effort than calculating
diagrams with the ordinary prescriptions \cite{UVQG,absograv}. Among the
first things to compute, we mention the one-loop self-energy diagrams, which
give the physical masses $\bar{m}$ and the widths $\Gamma $.

If $p^{2}-m^{2}$ is large enough, we can resum the bubble diagrams $B$ and
get the dressed fakeon propagators 
\begin{equation}
\mathbb{\bar{G}}_{\pm }=\mathbb{G}_{\pm }+\mathbb{G}_{\pm }B\mathbb{G}_{\pm
}+\mathbb{G}_{\pm }B\mathbb{G}_{\pm }B\mathbb{G}_{\pm }+\cdots =\frac{1}{%
\mathbb{G}_{\pm }^{-1}-B}.  \label{resum}
\end{equation}%
After the resummation, we can take $\mathcal{E}$ to zero, which gives%
\begin{equation*}
\mathbb{\bar{G}}_{\pm }\sim \pm \frac{Z}{p^{2}-\bar{m}^{2}+i\bar{m}\Gamma
_{\pm }}=\pm ZG_{+}(p,\bar{m},\bar{m}\Gamma _{\pm })
\end{equation*}%
around the physical peak $p^{2}=\bar{m}^{2}$, where $Z$ is the normalization
factor. The optical theorem implies 
\begin{equation*}
\mathrm{Im}[\mp ZG_{+}(p,\bar{m},\bar{m}\Gamma _{\pm })]=\frac{\bar{m}Z(\pm
\Gamma _{\pm })}{(p^{2}-\bar{m}^{2})^{2}+\bar{m}^{2}\Gamma _{\pm }^{2}}%
\geqslant 0,
\end{equation*}%
i.e. $\Gamma _{+}>0$, $\Gamma _{-}<0$: a fakeon plus has a positive width,
while a fakeon minus has a negative width. For $\Gamma _{\pm }\rightarrow
0^{\pm }$ we get%
\begin{equation}
\lim_{\Gamma _{\pm }\rightarrow 0^{\pm }}\mathrm{Im}[\mp ZG_{+}(p,\bar{m},%
\bar{m}\Gamma _{\pm })]\sim \pi Z\delta (p^{2}-\bar{m}^{2}).  \label{imma}
\end{equation}%
In the case of a physical particle, we would find exactly the same result,
which means that if we just watch the decay products of a fakeon, we have
the illusion of a true particle.

As said, the resummation (\ref{resum}) is legitimate only if $p^{2}-m^{2}$
is large enough. With physical particles, analyticity allows us to reach the
peak straightforwardly. However, fakeons just obey regionwise analyticity,
so we must be more careful. Indeed, the resummation misses the contact terms 
$\delta (p^{2}-m^{2})$, $\delta ^{\prime }(p^{2}-m^{2})$, etc. In general,
the sum of such contact terms plus (\ref{imma}) gives 
\begin{equation}
\sigma \pi Z\delta (p^{2}-\bar{m}^{2})  \label{immac}
\end{equation}%
for $\Gamma _{\pm }\rightarrow 0^{\pm }$, where $\sigma =1,0,-1$ in the case
of a physical particle, a fakeon and a ghost, respectively. Formula (\ref%
{immac}) tells us that if we try to detect the fakeons \textquotedblleft on
the fly\textquotedblright , we do not see anything. With a physical
particle, instead, what we obtain from the indirect observation, given by
formula (\ref{imma}), coincides with what we obtain from the direct
observation, given by formula (\ref{immac}). Finally, in the case of a
ghost, we have the illusion of a particle if we observe its decay products,
but get an absurdity (a \textquotedblleft minus one
particle\textquotedblright ), when we try and observe it on the fly.

The properties just outlined appear to justify the name \textquotedblleft
fakeon\textquotedblright , or fake particle. The fakeon can only be virtual,
so the only way to reveal\ it is by means of the interactions it mediates.

Since $\chi _{\mu \nu }$ is a fakeon minus, its width $\Gamma _{\chi }$ is
negative. In the case of the GFF theory, we find \cite{absograv} 
\begin{equation}
\Gamma _{\chi }=-C\frac{m_{\chi }^{3}}{M_{\mathrm{Pl}}^{2}},\qquad C=\frac{1%
}{120}(N_{s}+6N_{f}+12N_{v}),  \label{gamma}
\end{equation}%
where $N_{s}$, $N_{f}$ and $N_{v}$ are the numbers of (physical) scalars,
Dirac fermions (plus one half the number of Weyl fermions) and gauge
vectors, respectively. We are assuming that the masses of the matter fields
are much smaller than $m_{\chi }$, otherwise we have to include
mass-dependent corrections. Note that the graviton and the fakeons do not
contribute to $\Gamma _{\chi }$. In the GSF theory there is another
contribution due to $\phi $, which depends on $m_{\phi }$. The negative
width is a sign that microcausality is violated. However, it is not the only
way such a violation manifests itself, as we explain in the next section.

\section{Projection and classicization}

\label{class}

The generating functional $\Gamma (g_{\mu \nu },\phi ,\chi _{\mu \nu },\Phi
) $ of the one-particle irreducible correlation functions can be formally
projected by integrating out the fakeons, using the fakeon prescription.
This operation gives the physical $\Gamma $ functional. In some sense, the
fakeons can be viewed as auxiliary fields with kinetic terms.

For simplicity, consider an unprojected $\Gamma $ functional $\Gamma
(\varphi ,\chi )$, where $\varphi $ denotes the physical fields and $\chi $
denotes the fakeons. Solve the fakeon field equations $\delta \Gamma
(\varphi ,\chi )/\delta \chi =0$ by means of the fakeon prescription and
denote the solutions by $\langle \chi \rangle $. Then the physical, or
projected, $\Gamma $ functional $\Gamma _{\text{pr}}$ is%
\begin{equation*}
\Gamma _{\text{pr}}(\varphi )=\Gamma (\varphi ,\langle \chi \rangle ).
\end{equation*}%
Since the fakeons are not asymptotic states, at the level of the functional
integral it is sufficient to set their sources $J_{\chi }$ to zero:%
\begin{equation*}
Z_{\text{pr}}(J)=\int [\mathrm{d}\varphi \mathrm{d}\chi ]\exp \left(
iS(\varphi ,\chi )+i\int J\varphi \right) =\exp \left( iW_{\text{pr}%
}(J)\right) ,
\end{equation*}%
so $\Gamma _{\text{pr}}(\varphi )$ is the Legendre transform of $W_{\text{pr}%
}(J)$. Indeed, the unprojected formulas%
\begin{eqnarray*}
\Gamma (\varphi ,\chi ) &=&-W(J,J_{\chi })+\int J\varphi +\int J_{\chi }\chi
, \\
\varphi &=&\frac{\delta W(J,J_{\chi })}{\delta J},\qquad \chi =\frac{\delta
W(J,J_{\chi })}{\delta J_{\chi }},\qquad J=\frac{\delta \Gamma (\varphi
,\chi )}{\delta \varphi },\qquad J_{\chi }=\frac{\delta \Gamma (\varphi
,\chi )}{\delta \chi },
\end{eqnarray*}%
turn into the projected ones%
\begin{equation*}
\Gamma _{\text{pr}}(\varphi )=-W_{\text{pr}}(J)+\int J\varphi ,\qquad
\varphi =\frac{\delta W_{\text{pr}}(J)}{\delta J},\qquad J=\frac{\delta
\Gamma _{\text{pr}}(\varphi )}{\delta \varphi },
\end{equation*}%
when $J_{\chi }=0$.

In the classical limit, the fakeon prescription and the fakeon projection
simplify. In particular, the average continuation plays no role, because
there is no loop integral, so we can take (\ref{peps}) as it stands, which
gives the Cauchy principal value:%
\begin{equation}
\frac{p^{2}-m^{2}}{(p^{2}-m^{2})^{2}+\mathcal{E}^{4}}=\mathcal{P}\frac{1}{%
p^{2}-m^{2}}.  \label{fprc}
\end{equation}

To illustrate the projection in a simple case, consider the
higher-derivative Lagrangian 
\begin{equation}
\mathcal{L}_{\text{HD}}=\frac{m}{2}(\dot{x}^{2}-\tau ^{2}\ddot{x}^{2})+xF_{%
\text{ext}}(t),  \label{lagen2}
\end{equation}%
where $x$ is the coordinate, $m$ is the mass and $\tau $ is a real constant.
The unprojected equation of motion is 
\begin{equation*}
mK\ddot{x}=F_{\text{ext}},\qquad K=1+\tau ^{2}\frac{\mathrm{d}^{2}}{\mathrm{d%
}t^{2}},
\end{equation*}%
while the projected equation reads \cite{classicization} 
\begin{equation}
m\ddot{x}=\mathcal{P}\frac{1}{K}F_{\text{ext}}=\int_{-\infty }^{\infty }%
\mathrm{d}u\hspace{0in}\hspace{0.01in}\frac{\sin (|u|/\tau )}{2\tau }F_{%
\text{ext}}(t-u).  \label{average}
\end{equation}%
We see that the external force is convoluted with an oscillating function,
so the future ($u<0$) contributes as well as the past. This is how the
violation of microcausality survives the classical limit.

As for the classicization of quantum gravity in four dimensions, the
unprojected field equations derived from (\ref{SQG2}) are%
\begin{equation}
R^{\mu \nu }-\frac{1}{2}g^{\mu \nu }R=\frac{\kappa ^{2}}{\zeta }\left[ 
\mathrm{e}^{3\kappa \phi }fT_{\mathfrak{m}}^{\mu \nu }(\tilde{g}\mathrm{e}%
^{\kappa \phi },\Phi )+fT_{\phi }^{\mu \nu }(\tilde{g},\phi )+T_{\chi }^{\mu
\nu }(g,\chi )\right] ,  \label{mete}
\end{equation}%
for the metric tensor, and%
\begin{eqnarray}
-\frac{1}{\sqrt{-\tilde{g}}} &&\partial _{\mu }\left( \sqrt{-\tilde{g}}%
\tilde{g}^{\mu \nu }\partial _{\nu }\phi \right) -\frac{m_{\phi }^{2}}{%
\kappa }\left( \mathrm{e}^{\kappa \phi }-1\right) \mathrm{e}^{\kappa \phi }=%
\frac{\kappa \mathrm{e}^{3\kappa \phi }}{3\zeta }T_{\mathfrak{m}}^{\mu \nu }(%
\tilde{g}\mathrm{e}^{\kappa \phi },\Phi )\tilde{g}_{\mu \nu },  \notag \\
&&\frac{1}{\sqrt{-g}}\frac{\delta S_{\chi }(g,\chi )}{\delta \chi _{\mu \nu }%
}=\mathrm{e}^{3\kappa \phi }fT_{\mathfrak{m}}^{\mu \nu }(\tilde{g}\mathrm{e}%
^{\kappa \phi },\Phi )+fT_{\phi }^{\mu \nu }(\tilde{g},\phi ),  \label{fic}
\end{eqnarray}%
from the variations of $\phi $ and $\chi _{\mu \nu }$, where $T_{A}^{\mu \nu
}(g)=-(2/\sqrt{-g})(\delta S_{A}(g)/\delta g_{\mu \nu })$ are the
energy-momentum tensors ($A=\mathfrak{m}$, $\phi $, $\chi $) and $f=\sqrt{%
\det \tilde{g}_{\rho \sigma }/\det g_{\alpha \beta }}$.

The fakeon projection of the GSF\ theory is obtained by solving the second
line of (\ref{fic}) by means of the classical fakeon prescription, i.e. the
Cauchy principal value, and inserting the solution $\langle \chi _{\mu \nu
}\rangle $ into the other two equations. In the GFF theory, we have to solve
both equations (\ref{fic}) by means of the classical fakeon prescription and
insert the solutions $\langle \phi \rangle $, $\langle \chi _{\mu \nu
}\rangle $ into (\ref{mete}). The projected equations can also be obtained
from the finalized classical actions 
\begin{eqnarray}
\mathcal{S}_{\text{QG}}^{\text{GSF}}(g,\phi ,\Phi ) &=&S_{\text{H}%
}(g)+S_{\chi }(g,\langle \chi \rangle )+S_{\phi }(\bar{g},\phi )+S_{%
\mathfrak{m}}(\bar{g}\mathrm{e}^{\kappa \phi },\Phi ),  \notag \\
\mathcal{S}_{\text{QG}}^{\text{GFF}}(g,\Phi ) &=&S_{\text{H}}(g)+S_{\chi
}(g,\langle \chi \rangle )+S_{\phi }(\bar{g},\langle \phi \rangle )+S_{%
\mathfrak{m}}(\bar{g}\mathrm{e}^{\kappa \langle \phi \rangle },\Phi ),
\label{projeq}
\end{eqnarray}%
respectively, where~$\bar{g}_{\mu \nu }=g_{\mu \nu }+2\langle \chi _{\mu \nu
}\rangle $.

The interim, unprojected actions (\ref{SQG}) and (\ref{SQG2}) are local,
while the finalized actions (\ref{projeq})\ are nonlocal. These properties
remind us of the gauge-fixed actions, which are local, but unprojected, and
become nonlocal (with most types of gauge-fixing conditions), once the
Faddeev-Popov ghosts and the temporal and longitudinal components of the
gauge fields are projected away. However, there is an important difference
between the fakeon projection and the gauge projection, since the former
acts on the initial, final\ and intermediate states [$|a\rangle $, $%
|b\rangle $ and $|n\rangle $ in formula (\ref{V}), respectively], but not on
the virtual legs inside the diagrams, while the latter also acts on the
virtual legs. Thus, the gauge-trivial modes completely disappear, while the
fakeons leave an important remnant, which is the violation of causality at
energies larger than their masses.

The masses of the fakeons are free parameters. If their values are
sufficiently smaller that the Planck mass, we may be able to detect the
violation of microcausality in the foreseeable future. Moreover, formulas (%
\ref{projeq}) show that the violation of microcausality survives the
classical limit.

As said, the fakeon prescription is not classical, but emerges from the loop
corrections. The projected actions (\ref{projeq}) must be understood
perturbatively, since the parent quantum field theory that generates them is
formulated perturbatively. Thus, the classicization is also perturbative and
shares many features with the quantum theory it comes from, like the
impossibility to write down \textquotedblleft exact\textquotedblright\ field
equations and the important roles played by asymptotic series and
nonperturbative effects \cite{nonpertfakeon}. As far as we know, this
backlash of the quantization on the classical limit is unprecedented.

The nonrelativistic limit can be taken after the fakeon projection and,
possibly, the classicization. The fakeon propagator tends to the real part
of the usual quantum mechanical kernel. Note that both fakeons and
antifakeons contribute. For an analysis of nontrivial issues concerning the
nonrelativistic limit of quantum field theory, see ref. \cite{india}.

\section{The upgraded correspondence principle}

\setcounter{equation}{0}\label{upgra}

In this section we summarize the lessons learned from the previous ones in
connection with the correspondence principle and extend them to quantum
field theories of particles and fakeons in arbitrary spacetime dimensions.

\paragraph{\textbf{Unitarity}}

The unitarity requirement is unmodified, but better understood, since it
makes room for both particles and fakeons.

\paragraph{\textbf{Locality}}

The locality assumption must be upgraded, in the sense that it applies to
the interim classical action. The finalized classical action is generically
nonlocal, like the $S$ matrix and the generating functional $\Gamma $ of the
one-particle irreducible diagrams.

\paragraph{\textbf{Proper renormalizability}}

The renormalizability requirement, applied to the interim classical action,
must be formulated more precisely, since the usual notions are too generic.
We must demand \textit{proper} renormalizability, which is a refinement of
strict renormalizability. It means that the gauge couplings (including the
Newton constant) must be dimensionless (with respect to the power counting
governing the ultraviolet behaviors of the correlation functions), while the
other physical parameters must have nonnegative dimensions in units of mass.
The standard model does show that the gauge couplings have this particular
status among the couplings,\ so quantum gravity should conform to that.

\bigskip

We regard the three principles just listed as the cornerstones of the
correspondence principle of quantum field theory, and in particular quantum
gravity. If we remove the locality assumption, for example, we must guess
the $S$ matrix or the $\Gamma $ functional directly, which are infinitely
arbitrary. So doing, we have no way to determine the theory exhaustively,
since, as stressed in the introduction, when we explore the infinitesimal
world we cannot make infinitely many observations in a finite amount of time
and/or without disturbing the system. If we remove unitarity, we open the
way to the presence of ghosts, which leads to absurd behaviors. If we
renounce renormalizability, then we can just be satisfied with the
nonrenormalizable, low-energy theory of quantum gravity, obtained from the
Einstein-Hilbert Lagrangian plus the counterterms turned on by
renormalization \cite{qgravnew}.

In addition to the three basic requirements, we must include fundamental
symmetries, like Lorentz invariance, general covariance and gauge
invariance. Other properties are important, but not so much as to elevate
them to fundamental principles. Among those, we mention causality and
analyticity, which are downgraded to macrocausality and regionwise
analyticity, respectively.

\subsection{Uniqueness}

Is the resulting correspondence principle sufficient to point to a unique
theory? Various signals, like the arbitrariness of the matter sector of the
standard model, tell us that this might be a utopian goal. However, we do
have uniqueness in quantum gravity and a sort of uniqueness in form of the
gauge interactions.

Let us start from flat space. In every even spacetime dimensions $d\geqslant
4$ the correspondence principle made of unitarity, locality, proper
renormalizability and Lorentz invariance determines the gauge
transformations \cite{girard} and the form of the interim classical action,
which reads%
\begin{equation}
S_{\text{YM}}^{d}=-\frac{1}{4}\int \mathrm{d}^{d}x\sqrt{-g}\left[ F_{\mu \nu
}^{a}P_{(d-4)/2}(D^{2})F^{a\mu \nu }+\mathcal{O}(F^{3})\right] \text{,}
\label{YMD}
\end{equation}%
where $F_{\mu \nu }^{a}$ denotes the field strength, $D$ is the covariant
derivative, $P_{n}(x)$ is a real polynomial of degree $n$ in $x$ and $%
\mathcal{O}(F^{3})$ are the Lagrangian terms that have dimensions smaller
than or equal to $d$ and are built with at least three field strengths
and/or their covariant derivatives. The quadratic terms have been simplified
by means of Bianchi identities and partial integrations. As per proper
renormalizability, the gauge coupling is dimensionless. The coefficients of
the polynomial $P_{(d-4)/2}$ must satisfy suitable restrictions. In
particular, after projecting away the gauge modes, the massless poles of the
propagators must have positive residues and must be quantized as physical
particles. The other poles must have squared masses with nonnegative real
parts. The poles with negative or complex residues, as well as those with
positive residues but complex masses, must be quantized as fakeons. Finally,
the poles with positive residues and nonvanishing real masses can be
quantized either as fakeons or physical particles.

If we also demand microcausality, i.e. forbid the presence of fakeons, the
set of requirements implies that the spacetime dimension $d$ must be equal
to four. Then the action is the Yang-Mills one, 
\begin{equation}
S_{\text{YM}}=-\frac{1}{4}\int \mathrm{d}^{4}x\sqrt{-g}F_{\mu \nu
}^{a}F^{a\mu \nu }.  \label{SYM}
\end{equation}

Although the interim classical actions (\ref{YMD}) are essentially unique,
i.e. they contain finite numbers of independent parameters, we emphasize
that the gauge group remains free, as long as it is unitary and (together
with the matter content) satisfies the anomaly cancellation conditions
(which are other consequences of unitarity). In other words, the
correspondence principle fails to explain why the gauge group of the
standard model is the product of the three simplest groups, $U(1)$, $SU(2)$
and $SU(3)$, instead of anything else. For example, we cannot say why
factors such as $SU(13)$, $SU(19)$, etc., are absent.

It also fails to predict the matter content of the theory that describes
nature. Indeed, we are allowed to enlarge the standard model at will, to
include new massive particles and/or massive fakeons, as long as they are
heavy enough (to have no contradiction with experimental data) and the
anomaly cancellation conditions continue to hold. The ultimate theory of
nature could even contain infinitely many matter fields. In this respect,
the correspondence principle is almost completely powerless. So far, every
attempt (grand unification, supersymmetry, string theory and so on) to
relate the matter content to the interactions, beyond the anomaly
cancellation conditions, has failed. Probably, this is a sign of the fading
correspondence.

Nevertheless, quantum gravity turns out to be more unique than any other
theory. Indeed, its local symmetry (invariance under diffeomorphisms times
local Lorentz invariance) is unique and the requirements of unitarity,
locality, proper renormalizability and general covariance lead to the unique
interim classical actions (\ref{SQG})-(\ref{SQG2}) in four dimensions.

We also have solutions in even dimensions $d\geqslant 4$. Their interim
classical actions read%
\begin{equation}
S_{\text{QG}}^{d}=-\frac{1}{2\kappa ^{2}}\int \mathrm{d}^{d}x\sqrt{-g}\left[
2\Lambda _{C}+\zeta R+R_{\mu \nu }\mathcal{P}_{(d-4)/2}(D^{2})R^{\mu \nu }+R%
\mathcal{P}_{(d-4)/2}^{\prime }(D^{2})R+\mathcal{O}(R^{3})\right] ,
\label{SQGD}
\end{equation}%
where $\mathcal{P}_{n}$ and $\mathcal{P}_{n}^{\prime }$ denote other real
polynomials of degree $n$, while $\mathcal{O}(R^{3})$ are the Lagrangian
terms that have dimensions smaller than or equal to $d$, built with at least
three curvature tensors and/or their covariant derivatives. The free
propagators must satisfy the same requirements listed above and be quantized
as explained.

If we relax proper renormalizability into simple renormalizability, then we
lose most uniqueness properties, because there exist infinitely many
super-renormalizable theories of quantum gravity and gauge fields with
fakeons in every spacetime dimensions $d$, with interim actions equal to (%
\ref{YMD}) and (\ref{SQGD}), but polynomials $\mathcal{P}_{n}$ and $\mathcal{%
P}_{n}^{\prime }$ having degrees $n>(d-4)/2$.

Summarizing, the upgraded correspondence principle is made of 
\begin{equation}
\begin{tabular}{c}
$\text{unitarity}$ \\ 
$\text{locality}$ \\ 
$\text{proper renormalizability}$%
\end{tabular}
\label{corra}
\end{equation}%
together with fundamental symmetries and the requirements of having no
massless fakeons and finitely many fields and parameters. The combination (%
\ref{corra}) implies quantum gravity coupled to gauge and matter fields in
four dimensions, with interim classical actions (\ref{SQG})-(\ref{SQG2}).

With respect to the version of the correspondence principle that is
successful in flat space, the only upgrade required by quantum gravity
amounts to better understand the meanings of the principles themselves,
renounce analyticity in favor of regionwise analyticity and settle for
macrocausality instead of full causality. As we wanted at the beginning, the
final solution is as conservative as possible. The gravitational
interactions are essentially unique, the Yang-Mills interactions are unique
in form and the matter sector remains basically unrestricted.

\subsection{Causality}

Renouncing causality in quantum field theory is not a big sacrifice, because
we do not have a formulation that corresponds to the intuitive notion \cite%
{diagrammar}. What we have are off-shell formulations, such as Bogoliubov's
definition \cite{bogoliubov}, which also implies the
Lehmann-Symanzik-Zimmermann requirement that fields commute at spacelike
separated points \cite{LSZ}. The crucial issue is that it is not possible to
accurately localize spacetime points by working with relativistic wave
packets that correspond to particles that are on shell. This is more or less
the reason why microcausality has not been treated as a fundamental
principle in quantum field theory so far, maybe in anticipation that it was
going to be renounced eventually. We could even say that the fate of
causality was sealed from the birth of quantum field theory: quantum gravity
just delivered the killing blow. For a more detailed discussion on these
topics, see \cite{correspondence}.

\section{Conclusions}

\label{conclusions} \setcounter{equation}{0}

Various signals suggest that the correspondence between the macroscopic
environment\ where we live, which shapes our thinking, and the microscopic
world is doomed to become weaker and weaker as we explore smaller and
smaller distances. The impossibility to predict the gauge group and the
matter content of the theory of nature, as well as the fates of determinism
and causality are signs that our predictive power is fading away. We have to
cope with the fact that nature is not arranged to be understood or explained
by us humans to an arbitrary degree of precision. The ultimate theory of the
universe may look infinitely arbitrary to us. At the same time, the success
of quantum field theory and the recent progresses in quantum gravity give us
reasons to believe that we might still have a few interesting things to say
before declaring game over.

In this paper we have studied the properties of quantum field theory of
particles and fakeons in various dimensions. We have seen that the
correspondence principle that worked successfully for the standard model
admits a natural upgraded version that accommodates quantum gravity. It is
encoded in the requirements of unitarity, locality of the interim classical
action and proper renormalizability. The upgraded principle actually leads
to an essentially unique theory of quantum gravity in every even dimensions
greater than $2$. In four dimensions, a fakeon of spin 2 and a scalar field
are enough to have both unitarity and renormalizability. Causality breaks
down at energies larger than the fakeon masses. The classical limit shares
several features with the quantum theory it comes from, such as the
impossibility to write exact field equations.

Our experience teaches us that determinism and causality dominate at large
distances. On the other hand, when we explore smaller and smaller distances,
we see a gradual emergence of \textquotedblleft freedom\textquotedblright ,
first in the form of quantum uncertainty, then in the forms of acausality
and lack of time ordering. These facts suggest that the universe is radially
irreversible, i.e. irreversible in the sense of the relative distances. When
we move from the large to the small distances we see a pattern, pointing
from the absolute lack of freedom to what we may call asymptotic anarchy.

\end{document}